\documentclass[twocolumn,showpacs,preprintnumbers,nofootinbib]{revtex4-2}

\usepackage{graphicx}  
\usepackage{dcolumn}
\usepackage{bm}
\usepackage{amsmath,amssymb,amsfonts}
\usepackage{latexsym}
\usepackage{color}
\usepackage[colorlinks=true,urlcolor=black,anchorcolor=blue
,citecolor=blue,filecolor=blue,linkcolor=blue,menucolor=blue
,linktocpage=true,pdfproducer=medialab,pdfa=true]{hyperref}
\def\nn    {\nonumber}

\begin{document}
\title{\boldmath
Reconstructing the general 2HDM charged Higgs boson at the LHC}
\author{Wei-Shu Hou and Mohamed Krab}
\affiliation{Department of Physics, National Taiwan University, Taipei 10617, Taiwan}
	\bigskip
\begin{abstract}
We study the discovery prospects for a charged Higgs boson via 
the $b g\to c H^- \to c \bar t b$ process at the Large Hadron Collider (LHC).
Focusing on the general Two Higgs Doublet Model (G2HDM) that possesses
extra Yukawa couplings, the process is controlled by extra top couplings 
$\rho_{tc}$ and $\rho_{tt}$, which can drive electroweak baryogenesis (EWBG) 
to account for the baryon asymmetry of the Universe (BAU). We propose 
benchmark points (BPs) and demonstrate that evidence could emerge at 14 TeV LHC and luminosity of 300 fb$^{-1}$, with 
discovery potential at 600 fb$^{-1}$.
\end{abstract}
\maketitle

\section{Introduction} 
Particle physics is in an {\it impasse}: other than the $h(125)$ 
boson that is quite consistent with the Standard Model (SM) Higgs boson, no new physics 
(${\cal NNP}$) has emerged so far! We advocate G2HDM: two identical weak doublets, 
but allow a second set of Yukawa couplings aside from fermion masses. Though it has 
not gained much traction, it in fact has quite a few merits.

First, extra top Yukawa couplings $|\rho_{tt}|$, $|\rho_{tc}| \sim 1$ can each 
drive~\cite{Fuyuto:2017ewj} EWBG, while ${\mathcal O}(1)$ quartic couplings 
provide~\cite{Kanemura:2004ch} the prerequisite $1^{\rm st}$ order EW phase transition. 
Second, with CP violation (CPV) ${\mathcal O}(1)$ in strength as is needed for EWBG, 
one is vulnerable to precision tabletop electron EDM (eEDM) experiments, such as 
ACME~\cite{ACME:2018yjb} and JILA~\cite{Roussy:2022cmp}. However, a spectacular cancellation mechanism was uncovered~\cite{Fuyuto:2019svr}, rooted in the diagrammatics of 
two-loop diagrams, giving $|\rho_{ee}/\rho_{tt}| \sim \lambda_e/\lambda_t$, where the second 
ratio is nothing but $m_e/m_t$. Furthermore, one has a ``phase lock'', that $\arg\rho_{ee} =
 - \arg\rho_{tt}$, to cancel the ``$W$-loop Higgs-$\gamma$-$\gamma^*$'' insertion. Could this be 
the reason behind the ``flavor code'', that ${\mathcal Nature\;setup}$ fermion mass and mixing 
hierarchies as observed in SM couplings?
Third, the usual criticism~\cite{Glashow:1976nt} of G2HDM is {its possession of} flavor changing 
neutral couplings (FCNCs),~such as $t \to ch$~\cite{Hou:1991un}. Interestingly, to date we have 
not yet observed this plausible decay, as ${\mathcal Nature}$ seems to throw in small $h$-$H$ 
mixing (with $H$ the exotic CP-even boson), $c_\gamma \equiv \cos\gamma$ to control it --- {\it 
alignment}. ${\mathcal Nature}$ threw in a purely Higgs-sector parameter {\it to control FCNC!}
Fourth, small $c_\gamma$ does {\it not}~\cite{Hou:2017hiw} contradict ${\mathcal O}(1)$ 
quartics, e.g. $\eta_6$ in the second relation {of} Eq.~(2) below. Interestingly, one can then 
argue that the $H$, $A$ and $H^+$ bosons {\it could} populate 300--600~GeV. 
Finally, with $t \to ch$ $c_\gamma$-suppressed, sub-TeV exotic Higgs masses inspired the 
$cg \to tH/tA \to tt\bar c, tt\bar t$~\cite{Kohda:2017fkn} processes, which are unsuppressed 
by $s_\gamma \equiv \sin\gamma \simeq 1$; these were followed by the more
advantageous~\cite{Ghosh:2019exx} $cg \to bH^+ \to bt\bar b$ process, with a recoiling 
$b$-jet rather than a heavy top, and receiving Cabibbo-Kobayashi-Maskawa (CKM) enhancement 
compared to the popular SUSY type 2HDM-II.

\begin{figure}[b]
	\centering
	\includegraphics[width=.38 \textwidth]{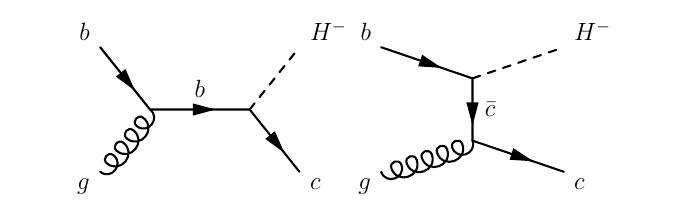}
	\caption{Feynman diagrams for the $bg \rightarrow c H^-$ process.}
	\label{diags}
\end{figure}

We investigate prospects for the $H^+$ boson in G2HDM to improve $H^+$ 
reconstruction compared to $cg \to bH^+ \to bt\bar b$~\cite{Ghosh:2019exx}. We suggest $pp \rightarrow cH^-$ (plus conjugate) search, which arises from the $bg \to cH^-$ 
parton process (Fig.~\ref{diags}), which is again {\it not} CKM-suppressed. The associated $c$-jet 
($\sim$ ``light quark'' jet) as tag-jet helps suppress background. We select five BPs that emulate Ref.~\cite{Ghosh:2019exx}, so $H^+ \to t\bar b$ decay is predominant, and 
present a signal-to-background analysis at 14~TeV LHC.

\vskip0.15cm

\section{G2HDM}
\subsection{Higgs Couplings}

With two identical scalar doublets, in the Higgs basis where only one doublet 
breaks the symmetry, the most general CP-invariant Higgs potential
is~\cite{Davidson:2005cw,Hou:2017hiw},
\begin{align}
 & V(\Phi,\Phi') = \mu_{11}^2|\Phi|^2 + \mu_{22}^2|\Phi'|^2
    - (\mu_{12}^2\Phi^\dagger\Phi' + \rm{h.c.}) \nn \\
 & \qquad + \frac{\eta_1}{2}|\Phi|^4 + \frac{\eta_2}{2}|\Phi'|^4
   + \eta_3|\Phi|^2|\Phi'|^2  + \eta_4 |\Phi^\dagger\Phi'|^2 \\
 & \qquad + \left[\frac{\eta_5}{2}(\Phi^\dagger\Phi')^2
   + \left(\eta_6 |\Phi|^2 + \eta_7|\Phi'|^2\right) \Phi^\dagger\Phi' + \rm{h.c.}\right], \nn
\end{align}
where $\eta_i$'s are quartic couplings and taken as real. $\Phi$ generates $\mathit{v}$ to break 
EW symmetry spontaneously via a first minimization condition, 
$\mu_{11}^2 = - \frac{1}{2}\eta_1 v^2$, while $\left\langle \Phi'\right\rangle = 0$ hence 
$\mu_{22}^2 > 0$. A second minimization condition, $\mu_{12}^2 = \frac{1}{2}\eta_6 v^2$, 
removes $\mu_{12}^2$ as a parameter.

Diagonalizing the $h$, $H$ mass-squared matrix gives~\cite{Davidson:2005cw,Hou:2017hiw}  mixing angle $\gamma$ ($\equiv \beta-\alpha$ in 2HDM-II notation),
\begin{align}
c^2_\gamma = \frac{\eta_1 v^2 - m_h^2}{m_H^2-m_h^2},
\quad s_\gamma c_\gamma = \frac{\eta_6 v^2}{m_H^2-m_h^2},
\end{align}
with approximate alignment implying $c_\gamma \simeq |\eta_6|v^2/(m^2_H - m^2_h)$~\cite{Hou:2017hiw}, as $s_\gamma$ is very close to 1.

The Higgs masses can be written in terms of the potential parameters in Eq.~(1),
\begin{align}
 & m_{H^+}^2 = \mu_{22}^2 +  \frac{1}{2}\eta_3 v^2, \ \;
     m_{A}^2 =  m^2_{H^+} + \frac{1}{2}(\eta_4 - \eta_5) v^2,\\
 & m_{H,h}^2 = \frac{1}{2}\bigg[m_A^2 + (\eta_1 + \eta_5) v^2\nn\\
 & \quad\quad \quad\quad \pm \sqrt{\left(m_A^2 + (\eta_5 - \eta_1) v^2\right)^2
   + 4 \eta_6^2 v^4}\,\bigg].	
\end{align}

The general Yukawa couplings are~\cite{Davidson:2005cw,Hou:2019mve}
\begin{align}
 \mathcal{L}_Y 
   = &\ \frac{1}{\sqrt{2}} \sum_{f = u, d, \ell}
           \bar f_{i} \bigg[\big(\lambda^f_{ij} s_\gamma - \rho^f_{ij} c_\gamma\big)\,h \nn \\
      & -\big(\lambda^f_{ij} c_\gamma + \rho^f_{ij} s_\gamma\big)H 
        +i\,{\rm sgn}(Q_f)\,\rho^f_{ij}\,A \bigg] R\,f_{j}\nn \\
      & -\bar{u}_i\big[(V\rho^d)_{ij} R - (\rho^{u\dagger}V)_{ij} L\big]d_j H^+ \nn \\
      & -\bar{\nu}_i\,\rho^\ell_{ij} R\,\ell_j H^+ +{\rm h.c.},
\end{align}
where $i,j = 1, 2, 3$ are generation indices, $L,R = (1\mp\gamma_5)/2$ and $V$ is the 
CKM matrix. The elements $\lambda^f_{ij} = \delta_{ij}\sqrt{2}m_i^f/v$ are real with 
$v \simeq 246~\rm{GeV}$, while $\rho^f_{ij}$ are non-diagonal and in general complex. 
Since we assume CP-conserving G2HDM, we shall take $\rho^f_{ij}$ as real in our study. 
We concentrate on $H^\pm$ production via $b g \rightarrow c H^-$ at the LHC 
(see Fig.~\ref{diags}), which is governed by the $\bar c bH^+$ vertex with coupling 
$\rho_{tc}V_{tb}$, as can be seen from Eq.~(5). 
We consider $H^+ \to t\bar b$ decay and study the $bg \to c H^- \to c \bar t b$ signal
(plus conjugate) at 14~TeV LHC. 

\subsection{Constrains on Parameter Space}
The parameter space is subject to various constraints. On theory side, we demand all parameters 
in Eq.~(1) to satisfy vacuum stability, tree-level unitarity and perturbativity, which are checked using the program \texttt{2HDMC-1.8.0}~\cite{Eriksson:2009ws}. As $\eta_{1}$, $\eta_{3-6}$ 
appear in exotic Higgs masses (Eqs.~(3) and (4)), we first express these quartic couplings in terms 
of $m^2_{h,H,A,H^+}$, $\mu^2_{22}$, $\gamma$, and $v$~\cite{Davidson:2005cw}, then randomly scan over 
$m_{H^+}$, $m_A$, $m_{H}$, $\mu^2_{22}$, $\eta_{2}$, $\eta_7$ within following ranges: 
$m_{H^+}\in [300, 600]$~GeV, $m_{H,A} \in [m_{H^+}-m_W, 650]$~GeV, 
$\mu^2_{22} \in [0, 10^6]$~GeV$^2$, $|\eta_{2,7}| \leq 3$ (this ${\cal O}(1)$ condition is 
imposed on all $\eta_i$s). We fix $m_h = 125$~GeV and set $c_\gamma = 0$, as $H^+$ couplings 
do not depend on $c_\gamma$. Thus, $\eta_1 = m^2_h/v^2 \cong 0.258$ and $\eta_6 = 0$.	

The scan is done via \texttt{2HDMC}, which employs $\Lambda_{1-7}$ and $m_{H^+}$ as 
Higgs basis inputs. We define $\eta_{1-7}$ as $\Lambda_{1-7}$, 
and require parameters to satisfy EW precision $S,\;T,\;U$~\cite{Peskin:1991sw} parameter 
constraints, with PDG fits (for $U = 0$)~\cite{ParticleDataGroup:2020ssz}: $S = 0.05 \pm 0.08$ 
and $T = 0.09 \pm 0.07$, with correlations taken into account. 

\begin{figure}[t]
\centering
\includegraphics[width=.3 \textwidth]{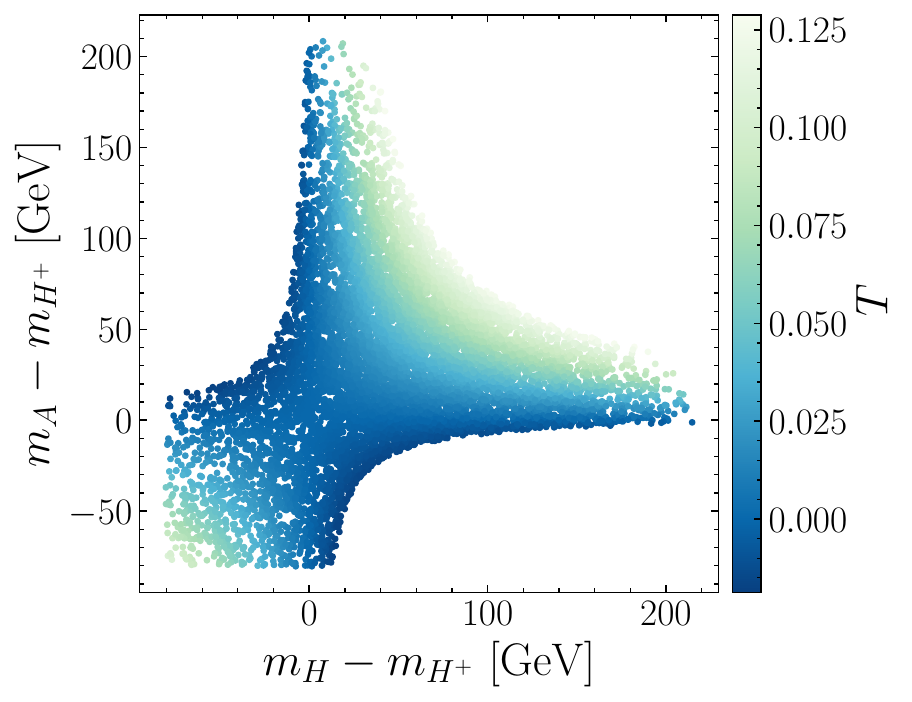}
\caption{Scan points for possible mass separation between $H^+$, $H$, and $A$ bosons. The color bar represents the $T$ parameter.}
\label{splitting}
\end{figure} 

The parameters satisfying theory and EW precision constraints (at $2\sigma$ level) are plotted 
in the ($m_H - m_{H^+}$, $m_A - m_{H^+}$) plane, as shown in Fig.~\ref{splitting}. 
The color code depicts the size of $T$ parameter, which constrains the masses
of $H$, $A$, $H^+$. We see that $m_{H^+}$ remains close to $m_H$ and/or $m_A$. Since 
we focus on $H^+ \to t\bar b$ decays as motivated by existing experimental $H^+$ searches, 
we assume the mass hierarchy $m_{H^+} \sim m_H$ and/or $m_{H^+} \sim m_A$. Thus, 
$H^+ \to W^+ H$ and/or $H^+ \to W^+ A$ decays are kinematically inaccessible 
(see Ref.~\cite{Hou:2021xiq} for implications of $H^+ \to W^+ A$ in G2HDM). 

We illustrate with five benchmarks, with $m_{H^+} =$ 300--500~GeV in 50~GeV steps, as listed in Table~\ref{table:BPs}. Note that BP1 and BP5 are BP1 and BP2 of Ref.~\cite{Ghosh:2019exx}. 

\begin{table}[b]
\centering
{\small  \begin{tabular}{c |c c c c  c  c c  c c} 
\hline
 BP & 
 $\eta_2$ & $\eta_3$ & $\eta_4$ & $\eta_5$ 
& $\eta_7$ & $m_{H^+}$ & $m_A$ & $m_H$ & ${\mu_{22}^2/v^2}$  \\
\hline
1 & 
1.40 & 0.62 & 0.53 & \,~ 1.06 & 
{$-0.79$} & 300 & 272 & 372 & 1.18 \\
2 & 
0.93 & 1.06 & 0.14 & {$-0.36$} & 
{$-0.22$} & 350 & 371 & 340 & 1.49 \\
3 & 
1.36 & 1.16 & 0.81 & \,~\,\,0.70 & 
{$-0.36$} & 400 & 404 & 454 & 2.06 \\
4 & 
0.61 & 1.83 & 1.30 & {$-0.30$} & 
\,~\,\,0.68 & 450 & 501 & 482 & 2.46 \\
5 & 
0.71 & 0.69 & 1.52 & {$-0.93$} & 
\,~\,\,0.24 & 500 & 569 & 517 & 3.78 \\
\hline
\end{tabular}}
\caption{Benchmark parameters for BP1-BP5. All masses in GeV,
with $\eta_6 = 0$, $m_h = 125$~GeV for all BPs.}
\label{table:BPs}
\end{table} 

There are experimental limits from flavor and collider physics. For simplicity, we set all 
$\rho_{ij} = 0$ other than the involved $\rho_{tc}$, $\rho_{tt}$ couplings. Flavor constraints 
are not stringent~\cite{Altunkaynak:2015twa, Crivellin:2013wna}. The constraints from 
$B_s$-$\bar{B}_s$ mixing and $b \to s\gamma$ on $\rho_{tc}$ are weak due to small 
$m_c$~\cite{Crivellin:2013wna}. It was found that $|\epsilon^u_{32}| \geq 1.3~(1.7)$ is excluded 
by $B_s$-$\bar{B}_s$ mixing for $m_{H^+} = 300\;(500)$~GeV and $\tan\beta = 50$. This leads to the bound $|\rho_{tc}| \lesssim 1.3~(1.7)$ for $m_{H^\pm} = 300\;(500)$~GeV.
For detailed discussion, see Ref.~\cite{Crivellin:2013wna}. 

The observables $B_q$-$\bar{B}_q$ ($q = s,d$) and $b \rightarrow s\gamma$ also put 
constraints on $\rho_{tt}$ and $\rho_{ct}$. The latter is strongly constrained by $B_{q}$-$\bar{B}_q$ due to enhanced CKM factor $|V_{cq}/V_{tq}| \sim 25$, hence $\rho_{ct}$ must be tiny \cite{Altunkaynak:2015twa}.
Regardless of $\rho_{ct}$, the limit on $\rho_{tt}$ is rather weak, leading to the upper bound $|\rho_{tt}| \lesssim 1.2\;(1.5)$, however, for $\rho_{ct} \lesssim 0.05\;(0.06)$, $0.5 \lesssim |\rho_{tt}| \lesssim 1.2\;(1.5)$ for $m_{H^+} = 300\;(500)$~GeV~\cite{Altunkaynak:2015twa}. Therefore, $\rho_{ct}$ is turned off.
Because of a $m_t/m_b$ enhancement factor, the limit from $b \rightarrow s\gamma$ constrains $\rho_{bb}$ more severely than $\rho_{tt}$, hence $\rho_{bb}$ is turned off. More details can be found 
in Ref.~\cite{Altunkaynak:2015twa}. Note that the selected benchmarks, 
$\rho_{tc},\,\rho_{tt} = 0.4, 0.6$, satisfy both constraints from $B_q$-$\bar{B}_q$ mixing 
and $b \to s\gamma$. 

\begin{figure}[t]
\centering
\includegraphics[width=.3 \textwidth]{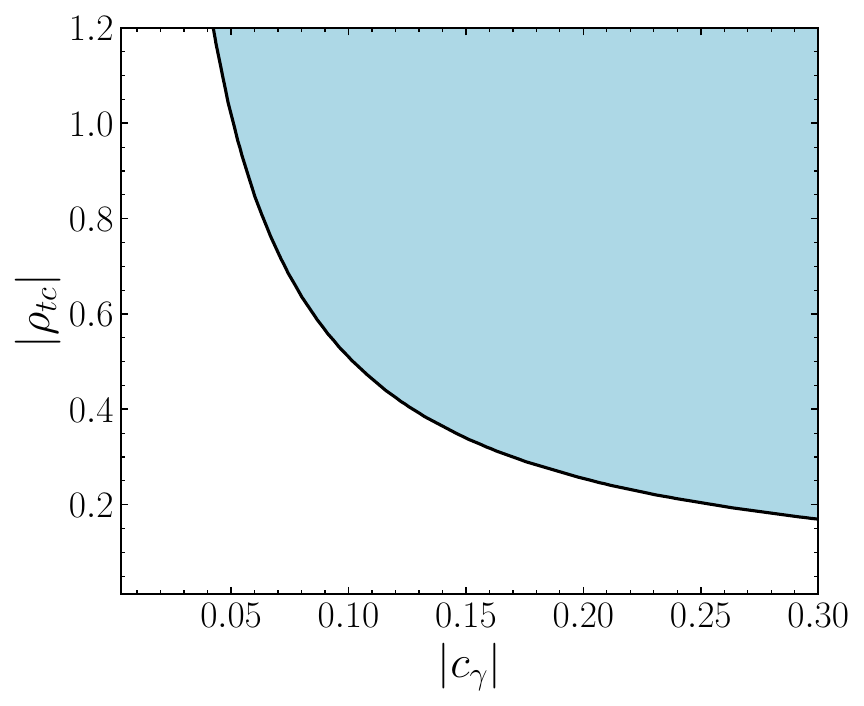}
\caption{ATLAS $t \rightarrow ch$ exclusion limit \cite{ATLAS:2024mih} in $|c_\gamma|$-$|\rho_{tc}|$ plane.}
\label{rhotc_limit}
\end{figure}

The $\rho_{tc}$, $\rho_{tt}$ couplings are further constrained by collider data. For 
$c_\gamma \neq 0$, $t \rightarrow ch$ searches set significant constraint on $\rho_{tc}$, where 
both CMS~\cite{CMS:2021hug} and ATLAS~\cite{ATLAS:2023ujo,ATLAS:2024mih} set  
95\% C.L. limits with full Run 2 data. We illustrate the most stringent ATLAS 
limit~\cite{ATLAS:2024mih} and find $|\rho_{tc}| \gtrsim 0.5$ is excluded at 95\% C.L. 
for $c_\gamma = 0.1$ (see Fig. \ref{rhotc_limit}). {The limit shrinks with $c_\gamma$ and 
disappears in alignment limit.} The $\rho_{tc}$ parameter receives further stringent constraint 
from CMS four top search~\cite{CMS:2019rvj}. 
See Refs.~\cite{Hou:2019gpn,Hou:2020chc,Hou:2021xiq} for more discussion.

\begin{figure}[t]
	\centering
	\includegraphics[width=.3 \textwidth]{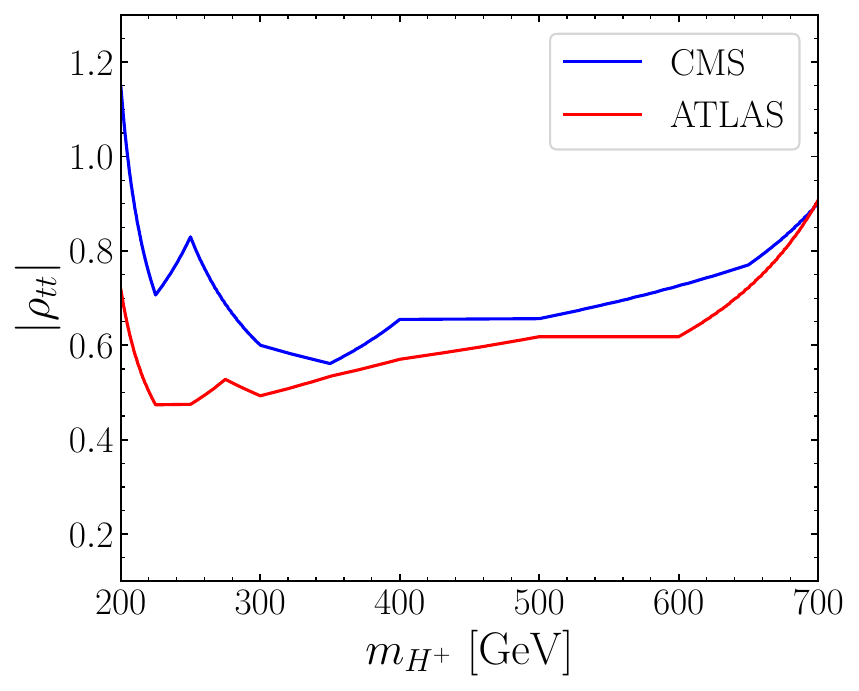}
	\caption{Exclusion bounds from ATLAS~\cite{ATLAS:2021upq} and CMS~\cite{CMS:2020imj}
		searches for $pp \rightarrow \bar t bH^+ \rightarrow \bar t bt\bar b$ in the 
		$m_{H^+}$-$\left|\rho_{tt}\right|$ plane.}
	\label{rhott_limit}
\end{figure}

ATLAS~\cite{ATLAS:2021upq} and CMS~\cite{CMS:2020imj} direct searches for 
$H^+ \to t\bar b$ at LHC strongly constrain $\rho_{tt}$. ATLAS uses full $139$~fb$^{-1}$ 
Run 2 data, while CMS used only $35.9$~fb$^{-1}$ so far, hence the ATLAS limit is more 
stringent on $\sigma(pp \to \bar t bH^+)\cdot{\cal B}(H^+ \rightarrow t\bar b)$ for $m_{H^+}$ 
between 0.2 to 2~TeV. We illustrate these limits assuming $\mathcal{B}(H^+ \to t\bar b) = 100\%$ 
to constrain $\rho_{tt}$ with leading order (LO) cross section estimates with 
\texttt{MadGraph5\_aMC@NLO}~\cite{Alwall:2014hca}, using the 2HDM model file in 
Ref.~\cite{Degrande:2014vpa}, and $K$-factor $\sim 1.6$~\cite{Degrande:2016hyf} to account 
for NLO corrections, as illustrated in Fig.~\ref{rhott_limit}; the CMS bounds are also depicted 
for comparison. LHC searches for $pp \to H/A \to t\bar t$~\cite{ATLAS:2017snw,CMS:2019pzc} 
and $pp \to ttH/A \to t\bar tt\bar t$~\cite{CMS:2019rvj} also constrain $\rho_{tt}$, but these
constraints are slightly weaker than direct $H^+$ searches and $B_{d,s}$ 
mixing~\cite{Ghosh:2019exx}. Direct and indirect LHC measurements can put further bounds 
on $\rho_{tt}$ (also $c_\gamma$~\cite{Hou:2018uvr}), specifically $t\bar th$ and Higgs property 
measurements. These bounds suffer $c_\gamma$ suppression, however, as seen from Eq.~(5). 
Note that our chosen $\rho_{tt}$ is safe from all constraints mentioned.

\section{Collider Study}

We study $H^-$ production in association with a $c$ quark, $bg \rightarrow cH^-$ (see Fig.~\ref{diags}). With $H^- \to \bar tb$ decay for all BPs, the signal would have three jets,
at least two identified as $b$-jets, plus one lepton and missing transverse momentum. 
With this novel signature, a $b$-jet and the lepton plus neutrino can be used to reconstruct 
a top, then combine with the other $b$-jet to reconstruct the $H^+$. But for the $bt\bar b$ 
signature of Ref.~\cite{Ghosh:2019exx} with three $b$-jets (or even worse for $\bar t bH^+ \to \bar tbt\bar b$), 
the high $b$-jet multiplicity makes $H^+$ reconstruction more difficult. Our final state has a 
further high $p_T$ ``tag''-jet. Thus, our signature is complementary to existing direct searches 
for $H^+$.

We follow the $cg \rightarrow \bar bH^+ \rightarrow b t\bar b$ analysis of 
Ref.~\cite{Ghosh:2019exx}, which has one extra $b$-tagged jet.\footnote{Our signal is included 
as subdominant contribution to the one proposed in Ref.~\cite{Ghosh:2019exx}, but killed by applied cuts.} {The analysis was extended further to improve sensitivity~\cite{Desai:2022zig}.} 

%

For the BPs listed in Table~\ref{table:BPs}, for $\rho_{tc} = 0.4$, $\rho_{tt} = 0.6$, 
the ${\cal B}(H^+\;\to\;c\bar b)$, ${\cal B}(H^+\;\to\;t\bar b)$ values are
50\% (44\%, 40\%, 38\%, 36\%) and 50\%~(56\%, 60\%, 62\%, 64\%) for 
BP1~(BP2, BP3, BP4, BP5), respectively.\footnote{
$H^+ \to c\bar b$ decay dominates at low $m_{H^+}$, especially below $m_t$, where the 
$H^+ \to t\bar b$ decay is kinematically forbidden.} Considering the $bg \to c H^- \to c \bar tb$ 
signal, assuming $t \to \ell$ {($e$ or $\mu$)} + $\nu$ + $b$-jet decay, the final state should 
be {two $b$-jets, one high $p_T$ jet, plus one lepton and missing $E_T$ from $\nu$.} 
The subdominant $bg \to tH^- \to t\bar cb$ is also included as signal. The main background is 
SM $t\bar t$ production in association with flavor jets. 
Other backgrounds are single top ($tj$), $Wt + \rm{jets}$, $t\bar th$ and $t\bar t Z$. 
Drell-Yan, $W+\rm{jets}$, $t\bar tW$ and $tWh$ backgrounds are minor.

Signal and background cross sections are computed at LO using~\texttt{MadGraph5\_aMC@NLO} 
with default \texttt{NN23LO1}~PDF~at~14~TeV~collision~energy.~All samples are passed 
through \texttt{Pythia8}\;\cite{Sjostrand:2014zea} for parton showering and hadronization, then 
processed through fast detector simulator \texttt{Delphes}\;\cite{deFavereau:2013fsa} with ATLAS 
card and anti-kt jet algorithm~\cite{Cacciari:2008gp}, and with $\Delta R =0.5$. The resulting 
signal and background events are analyzed using \texttt{MadAnalysis5}~\cite{Conte:2012fm}.
Backgrounds are rescaled using $K$-factors to account for NLO (or higher) QCD corrections. 
The $K$-factors for $t\bar t + \rm{jets}$, $Wt+\rm{jets}$, $t$($s$)-channel single top, 
$t\bar th$ and $t\bar{t} Z$ processes are 1.84~\cite{twiki}, 1.35~\cite{Kidonakis:2010ux}, 
1.2 (1.47)~\cite{twikisingtop}, 1.27~\cite{twikittbarh} and 1.56~\cite{Campbell:2013yla}, 
respectively. Signal cross sections are kept at LO.

\begin{table}[b]
\centering
{\small
\begin{tabular}{c |c c c c  c  c } 
\hline
 BP & $t\bar t + 2j$ & $Wt + 2j$ & $tj+1j$ & $t\bar th$ & $t\bar tZ$
 & $\rm{Signal}$  \\
\hline
 1 & 3143.2 & 699.2 & 228.1 & 1.5 & 0.9 & 14.9 \\
 2 & 2237.9 & 548.3 & 185.8 & 1.4 & 0.8 & 11.9 \\
 3 & 2782.1 & 816.5 & 222.5 & 1.9 & 1.1 & 13.8 \\
 4 & 2438.2 & 752.2 & 157.5 & 2.0 & 1.4 & 9.8 \\
 5 & 1894.8 & 605.5 & 108.1 & 1.7 & 1.0 & 6.4 \\
\hline
\end{tabular}}
\caption{
Cross sections (fb) at $14$ TeV after selection cuts.
}
\label{table:SBxs}
\end{table}

Candidate signal events are with at least two jets and no more than four jets, at least two 
$b$-tagged with $p^{j}_T > 20$~GeV, plus one lepton with $p^{\ell}_T > 30$~GeV, and 
$E^{\rm{miss}}_T > 20$~GeV. The angular separation $\Delta R$ between all jet-pairs, and any jet plus lepton should be larger than 0.4. The pseudo-rapidity $|\eta|$ of lepton and all jets should satisfy $|\eta| < 2.5$. The $H_T$ sum of lepton, leading jet and two $b$-jets should be 
larger than 350 (400)~GeV for BP1 (BP2-5). To further reduce backgrounds, especially 
$t\bar t + \rm{jets}$, $Wt+\rm{jets}$ and $tj$ contributions, the transverse mass of reconstructed 
$H^+$ should lie within $m_{H^+} \pm 50$~GeV mass window. We give the background 
and signal cross sections in Table~\ref{table:SBxs} for each BP. 

\begin{figure}[t!]
	\centering
	\includegraphics[width=.3 \textwidth]{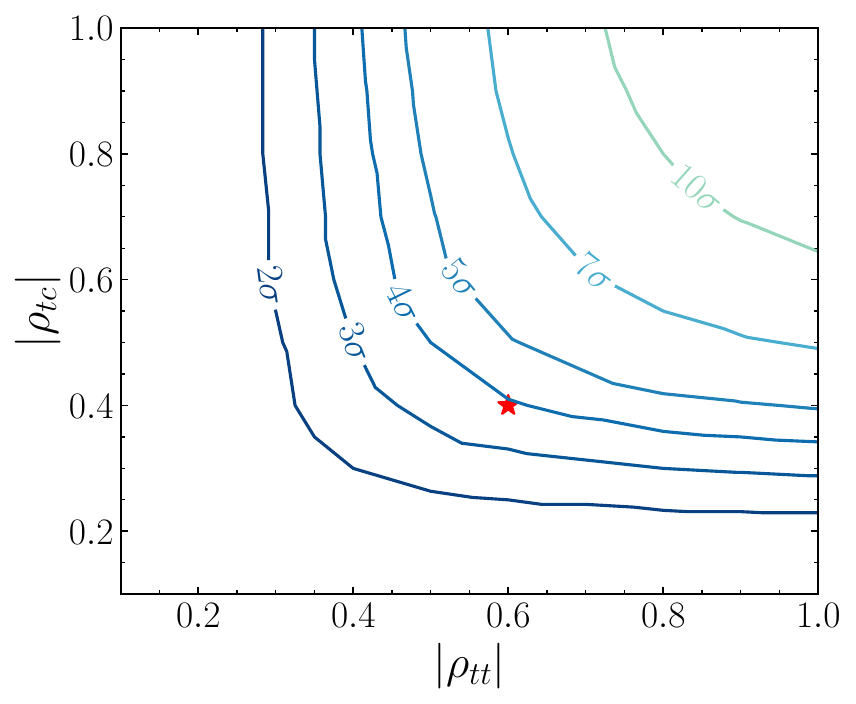}
	\caption{Significance in the $\left|\rho_{tt}\right|$-$\left|\rho_{tc}\right|$ plane. BP4, which corresponds to a significance of $\sim 3.9\sigma$, is shown as red star.}
	\label{cont:plot}
\end{figure}

We estimate our signal sensitivity using 
$\mathcal{Z} = \sqrt{2\left[(S+B)\ln(1+S/B)-S\right]}$~\cite{Cowan:2010js} for statistical 
significance, where $S$ is number of signal events and $B$ for background events. 
We find significance for BP1 (BP2, BP3) $\simeq 4.0\sigma$ ($3.8\sigma$, $3.9\sigma$) at 300~fb$^{-1}$, and $\simeq 5.7\sigma$ ($5.3\sigma$, $5.5\sigma$)
at 600~fb$^{-1}$. 
With 140~fb$^{-1}$, one has $\sim 2.8\sigma$ ($2.6\sigma$, $2.6\sigma$) significance for BP1 (BP2, BP3).
Thus, evidence (hint) could emerge with 300 (140)~fb$^{-1}$ data, while $600$~fb$^{-1}$ 
could claim discovery. 
With $m_{H^+}=400~\rm{GeV}$ (BP3) and $L=300~\rm{fb}^{-1}$, $\rho_{tc}=0.3$ and $\rho_{tt}=0.4$ can give $\mathcal{Z} \simeq 2.0\sigma$, while larger values could yield  $\mathcal{Z} \sim 5\sigma$ or higher (see Fig. \ref{cont:plot}).
For BP4 and BP5, we find $\simeq 2.9\sigma$ ($4.1$) and $\sim 2.2\sigma$ ($3.1\sigma$) at 300~fb$^{-1}$ (600~fb$^{-1}$), respectively. The significance at 140 fb$^{-1}$ is $\simeq 2.0\sigma$ for BP4 and less than $2\sigma$ for BP5. 
For these BPs, evidence (hint) should emerge at 600 (300)~fb$^{-1}$.
Note that the significance for $cg \to bH^+ \to bt\bar b$ is $\simeq$ 4.9 (5.0) and $\simeq$ 6.9 (7.1) for BP1 (BP5) at 300 and 600 fb$^{-1}$, respectively \cite{Ghosh:2019exx}.

Before closing, we note the $\rho_{tu}$ coupling can induce $bg \to uH^- \to ut\bar b$, but 
$\rho_{tu}$ is highly constrained. For $\rho_{tt} = 0.6$, it was found that~\cite{Ghosh:2019exx} with all other $\rho_{ij} = 0$, $\left|\rho_{tu} \right| \gtrsim 0.1~(0.2)$ is excluded at 95\% C.L. 
by CRW of Ref.~\cite{CMS:2019rvj} for BP1 (BP5). For instance, taking $\rho_{tu} = 0.1$ and 
$\rho_{tt} = 0.6$ (setting $\rho_{tc} = 0$), ${\cal B}(H^+ \to u\bar{b})$ and 
${\cal B}(H^+ \to t\bar{b})$ are 6\% and 94\%, respectively. The achievable significance 
for BP1 is less than $1\sigma$ even for 600 fb$^{-1}$. We therefore neglect the $\rho_{tu}$ 
contribution in our study.

In addition, the presence of other $\rho_{ij}$s, e.g. $\rho_{\tau\tau}$ would induce $H^+ \to \tau^+ \nu_\tau$ decay, which can dilute ${\cal B}(H^+ \to t\bar b)$ hence 
the signal. Taking $\rho_{\tau\tau} \sim \lambda_\tau$, however, would not change our 
conclusions because ${\cal B}(H^+ \to \tau^+ \nu_\tau)$ is tiny. Non-zero $\rho_{bb}$ 
induces $H^+ \to t\bar b$ decay and hence can yield $\bar cH^+$ signature. But for 
$\rho_{bb} \sim \lambda_b$, the contribution is negligible compared to $\rho_{tt} \sim 0.6$.

%
%

\section{Discussion and Summary}
Searches for $H^+$ boson with mass above $m_t$ (typically called ``heavy'' $H^+$) have 
relied on associated production with a top and bottom quark in the 4-flavor scheme, 
$pp \rightarrow \bar t bH^+$, and with top quark in the 5-flavor scheme, $pp \rightarrow
 \bar t H^+$. These processes were mostly motivated by MSSM (and 2HDM-II). In G2HDM, 
the novel $pp \rightarrow bH^+$ process (dominated by $cg$-initiated channel due to 
$\rho_{tc}V_{tb}$ coupling) was proposed~\cite{Ghosh:2019exx}. Assuming $H^+ \to t\bar b$ 
decay, it would yield a signature with at least three $b$-jets, one lepton and missing $p_T$. 
The high $b$-jet multiplicity makes it relatively difficult to reconstruct $H^+$. 

We propose a search for a charged Higgs boson in association with a light quark jet, with 
production cross section larger than typical $H^+$ production in association with a top quark, 
with $H^+ \to t\bar b$ decay. This novel signal would have two $b$-jets, one extra jet plus one 
lepton and missing $E_T$. It would be useful to probe $H^+$ further at the LHC with this alternative signature, {\it especially} if one sees a hint in $pp \to bH^+ \to bt\bar b$.
The proposed signal can be useful to discriminate G2HDM from other two Higgs doublet 
extensions, such as 2HDM-II, where the $bg \rightarrow cH^-$ process is CKM suppressed. 

The presence of the FCNC coupling $\rho_{tc}$ has significant impact on the $Z_2$ 
forbidden tree-level Higgs production process like 
$pp \rightarrow tH/A$~\cite{Kohda:2017fkn,Hou:2018zmg} (as well as controlling $V_{tb}$ proportional $pp \rightarrow H^+ \bar c/b$ processes)
and decay channels like $t \rightarrow ch$~\cite{Hou:1991un}, and 
$H/A \rightarrow t\bar c$~\cite{Hou:2018zmg}. These provide substantial avenues to 
complement existing direct and indirect LHC searches for exotic Higgs bosons (also at low energy). 
Moreover, $\rho_{tc}$ can drive electroweak baryogenesis in case $\rho_{tt}$ becomes 
ineffective~\cite{Fuyuto:2017ewj,Fuyuto:2019svr}. Thus, $\rho_{tc}$ can crucially open 
the door for more intriguing new physics scenarios. 
Observing those induced $\rho_{tc}$ processes would be a smoking gun for G2HDM. 

In summary, we investigate the discovery prospect for a charged Higgs boson through 
the $bg \rightarrow qH^- \rightarrow q\bar tb$ process. We present five BPs, with $H^+$ masses 
ranging from 300 to 500 GeV, and did a signal-to-background study. We find evidence could 
emerge with 300~ fb$^{-1}$ at 14 TeV collision, and possible discovery at $600$~fb$^{-1}$. 

\vskip0.2cm
\noindent{\bf Acknowledgments.--} 
This work is supported by NSTC 112-2639-M-002-006-ASP of Taiwan,
and NTU 113L86001 and 113L891801.


\end{document}